# Position Modulation Code for Rewriting Write-Once Memories

Yunnan Wu and Anxiao (Andrew) Jiang

*Abstract*—A write-once memory (wom) is a storage medium formed by a number of "write-once" bit positions (wits), where each wit initially is in a '0' state and can be changed to a '1' state irreversibly. Examples of write-once memories include SLC flash memories and optical disks. This paper presents a low complexity coding scheme for rewriting such write-once memories, which is applicable to general problem configurations. The proposed scheme is called the *position modulation code*, as it uses the positions of the zero symbols to encode some information. The proposed technique can achieve code rates higher than state-of-the-art practical solutions for some configurations. For instance, there is a position modulation code that can write 56 bits 10 times on 278 wits, achieving rate 2.01. In addition, the position modulation code is shown to achieve a rate at least half of the optimal rate.

## I. INTRODUCTION

In their pioneering work [18], Rivest and Shamir considered the problem of rewriting a "write-once memory" (wom). A write-once memory consists of a number of "write-once" cells, where each wit initially is in a '0' state and can be changed to a '1' state irreversibly. Several types of storage media follow such a write-once memory model. Examples include SLC flash memories and optical disks. Rivest and Shamir [18] demonstrated that it is possible to rewrite such a write-once memory multiple times, using coding techniques. For example, using 3 wits, we can write a 2-bit variable twice via a wom-code. The *rate* of this wom-code (i.e., capacity per wit) is $2 \times 2/3 = 1.33$.

In [18], Rivest and Shamir presented a wom-code that has the best asymptotic rate (as the cardinality of the variable approaches infinity), for any given number of writes. This result is done via a counting argument and the scheme is not constructive. Rivest and Shamir and others have proposed various other wom-codes, which have low complexity. However, the coding rates of existing wom-codes with low complexity are still far from the theoretically achievable coding rates. Thus, the problem of finding good practical wom-codes remains an open challenge.

In this paper, we present a low complexity coding scheme, which we call the *position modulation code*, for rewriting a write-once memory. The scheme is built upon a simple yet fundamental observation: If we flip $k$ cells out of $n$ binary cells that are initially zero, there are $n$-choose-$k$ ways of

Yunnan Wu is with Microsoft Research, One Microsoft Way, Redmond, WA, 98052. yunnanwu@microsoft.com. Anxiao (Andrew) Jiang is with the Computer Science and Engineering Dept., Texas A&M University, College Station, TX 77843. ajiang@cse.tamu.edu

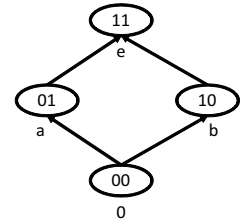

Fig. 1. Two wits are used to implement a symbol that can take four values $\{0, a, b, e\}$. The all-one value is used to represent the "erased" state.

doing so, which can represent $n$-choose-$k$ distinct messages. We call this encoding method *position modulation*. Clearly, position modulation can be directly applied to the first write in write-once memories, because we start from the all-zero state. Several wom-codes can be viewed as performing position modulation in the first write. However, it is not straightforward to apply position modulation to subsequent writes, because each write flips some wits, and the decoder only sees the overall set of written wits but not when they are written.

The scheme proposed in this paper performs position modulation to all writes except the last one. Instead of directly applying position modulation to the wits, we apply position modulation to symbols that are formed by multiple wits each. For example, as illustrated in Figure 1, using every two wits in a group, we obtain a symbol that can take four values. The all-one value is used to represent the "erased" state. At the beginning of each write, we erase all nonzero symbols by setting them to the all-one value. After this operation, the remaining symbols are all zero and thus we can apply position modulation.

The rest of the paper is organized as follows. In Section II, we present a brief overview of existing research on wom coding and related topics. In Section III, we highlight a simple observation that encoding and decoding for position modulation can be implemented with polynomial time complexity. The proposed position modulation code is then described in Section IV. In Section V we compare the performance of the position modulation code with existing wom-codes. In Section VI we present the conclusions.

## II. OVERVIEW OF RELATED STUDIES

We briefly overview the existing studies on wom codes. They will be referred to with more details later in the paper when we compare the performance of different codes. Rivest and Shamir defined wom codes in their pioneering paper [18], in which they presented a number of individual wom codes

(the most well-known of which is the code for writing two bits twice in three wits) and some families of wom codes (including the linear wom codes and the tabular codes). The wom code model was generalized in [4] by allowing the cell state transitions to be any acyclic directed graph; in addition, two families of wom codes for data with alphabet size three and four were shown. In [16] several individual wom codes were constructed using projective geometries. In [3] a wom code construction based on error-correcting codes was presented. As an example, based on the $C[23, 12, 7]$ Golay code, a wom code for writing 11 bits three times in 23 wits was shown. More works on wom include [6], [9], [21], [22], which studied the capacity and error correction of write-once memories.

WOM is related to the study on defective memory [8], [10], [14], write unidirectional memory [17], [19], [20], and write efficient memory [1], [7]. It is also related to the study on coding for flash memories, where many of the proposed coding schemes are based on the monotonic transitions of flash cell states [11], [13]. In particular, the works on rewriting codes for flash memories extend wom codes [2], [5], [12], [15].

The motivation for this study is to look for a general method for constructing wom codes that can achieve low encoding/decoding complexity and high rates, which can potentially be used in practice (e.g., for flash memories). As a result, our focus is on cases where the number of rewrites is reasonably small and the data size is moderate, instead of asymptotic settings.

## III. POSITION MODULATION HAS POLYNOMIAL COMPLEXITY

We use the following notations for wom codes in [18]. A $\langle v \rangle^t / n$ womcode is a code that can write a variable of cardinality $v$ $t$ times, using $n$ wits. More generally, a $\langle v_1, \ldots, v_t \rangle / n$ womcode is a code that can write a variable of cardinality $v_1$ the first time, a variable of cardinality $v_2$ the second time, and so on, using $n$ wits. The $i$-th write is also called the $i$-th generation. The rate of the code is:

$$\frac{\log_2(v_1 \ldots v_t)}{n}. \quad (1)$$

In position modulation, we use a length-$n$ binary vector with $k$ ones to represent a variable of cardinality $\binom{n}{k}$. Let $\mathcal{U}$ denote the set of all length-$n$ binary vectors with $k$ ones. The most natural approach is to associate each vector $\boldsymbol{u} \in \mathcal{U}$ with its index in the lexical sorted list of $\mathcal{U}$. Let

$$\ell : \mathcal{U} \mapsto \left\{0, 1, \ldots, \binom{n}{k} - 1\right\} \quad (2)$$

be a function that computes the lexical order of a vector $\boldsymbol{u} \in \mathcal{U}$. We now show a simple observation, that the encoding and decoding functions $\ell(\cdot)$ and $\ell^{-1}(\cdot)$ can be implemented efficiently.

First, consider an example where $n = 7$ and $k = 3$. Figure 2 illustrates the process of computing the lexical order of the sequence 0101100. The idea is to count the number of sequences that have a lower index. We start with the leftmost

$$\begin{array}{ll} \texttt{00*****} & \binom{5}{3} \\ \texttt{0100***} & \binom{3}{2} \\ \texttt{01010**} & \binom{2}{1} \end{array}$$

Fig. 2. Illustration of the process of computing the lexical order of the sequence 0101100.

1 and move to the right. First, if we flip the leftmost 1 to 0, then clearly vectors in $\mathcal{U}$ of the form $00****$ are all lexically before 0101100; there are $\binom{5}{3}$ of such vectors. If we keep the second bit as 0, the next group of vectors with a lower index than 0101100 is of the form $0100***$. There are $\binom{3}{2}$ of such vectors. If we keep the first two 1's, the next group of vectors with a lower index than 0101100 is of the form $01010**$. There are $\binom{2}{1}$ of such vectors. Therefore, the index of 0101100 is

$$g(0101100) = \binom{5}{3} + \binom{3}{2} + \binom{2}{1} = 15.$$

More generally, consider a vector $\boldsymbol{u} \in \boldsymbol{U}$ that has $k$ one's at positions $n > i_1 > \ldots > i_k \geq 0$ (here the bit positions are labeled as 0 to $n-1$ from right to left). We have:

$$\ell(\boldsymbol{u}) = \binom{i_1}{k} + \binom{i_2}{k-1} + \ldots + \binom{i_k}{1}. \quad (3)$$

Conversely, given $\ell(\boldsymbol{u})$, to find $\boldsymbol{u}$, we determine the bits from left to right. First, $i_1$ is determined as the largest integer such that $\binom{i_1}{k} \leq \ell(\boldsymbol{u})$. Next, $i_2$ is determined as the largest integer such that $\binom{i_2}{k-1} \leq \ell(\boldsymbol{u}) - \binom{i_1}{k}$. This process can be continued until all ones have been determined. All the above computation has time complexity polynomial in $n$.

## IV. POSITION MODULATION CODE

**Theorem 1:** Given integers $v_1, \ldots, v_t$ and $m \geq 2$, let $h_i$, $1 \leq i \leq t$ be integers satisfying

(a) $h_1 > h_2 > \ldots > h_t > 0$.
(b) $\sum_{k=0}^{h_1-h_2} \binom{h_1}{k} (2^m - 1)^k \geq v_1$,
(c) $\sum_{k=1}^{h_i-h_{i+1}} \binom{h_i}{k} (2^m - 2)^k \geq v_i, \quad i = 2, \ldots, t-1.$
(d) $(2^m - 1)^{h_t} - 1 \geq v_t$.

Then, there exists a $\langle v_1, \ldots, v_t \rangle / (m \times h_1)$ wom-code.
**Proof:** We organize the $n = mh_1$ wits as $h_1$ groups, where each group consists of $m$ wits and represents a symbol that can take $2^m$ values. The state of the storage cells is then described by a vector $\boldsymbol{x} = [x_1, \ldots, x_{h_1}]$, where each component can take values from $\{0, 1, \ldots, 2^m - 1\}$ (each value corresponds to its binary representation). Two of the states are special, the zero state implemented by the all-zero codeword and the erased state implemented by the all-one codeword.

At the beginning of each write, we erase all nonzero symbols by setting them to the all-one value. The remaining symbols are thus all zeros. Using these remaining symbols, we encode the message by the number of zero symbols, the positions of the zero symbols, and the values of the nonzero symbols.

We now describe the encoding process in detail. For the first write, we select 0 up to $h_1 - h_2$ symbols and write a variable with value from $\{1, \ldots, 2^m - 1\}$ to each of them. If we write into $k$ symbols, then the positions of the $k$ symbols can represent a variable of cardinality $\binom{h_1}{k}$, and the values of these $k$ symbols can represent a variable of cardinality $(2^m - 1)^k$. Since we can write $0, \ldots, h_1 - h_2$ symbols, in total we can represent a variable of cardinality

$$\sum_{k=0}^{h_1 - h_2} \binom{h_1}{k} (2^m - 1)^k, \tag{4}$$

which is at least $v_1$ according to condition (b). Hence the first write can be done.

For the $i$-th write with $1 < i < t$, first we erase all nonzero symbols by changing them into the all-one state. If there are more than $h_i$ remaining zero symbols, then we arbitrarily erase some additional zero symbols so that there are exactly $h_i$ remaining zeros. Then among the $h_i$ remaining zero symbols, we pick 1 up to $h_i - h_{i+1}$ entries and write a value from $\{1, \ldots, 2^m - 2\}$ to each of them. Since we can write $1, \ldots, h_i - h_{i+1}$ symbols, in total we can represent a variable of cardinality

$$\sum_{k=1}^{h_i - h_{i+1}} \binom{h_i}{k} (2^m - 2)^k \tag{5}$$

which is at least $v_i$ according to condition (c). Hence the $i$-th write can be done.

The last write is done differently in that position modulation is not used. First we erase all nonzero symbols and possibly some additional zero symbols so that there are exactly $h_t$ zero symbols. We simply use these $h_t$ positions to represent the variable by setting each position from $\{0, 1, \ldots, 2^m - 2\}$, except that the all zero codeword is not used. Since $(2^m - 1)^{h_t} - 1 \geq v_t$, the last write can be done.

Decoding is done accordingly. First, the generation number (i.e., which write) is decoded from the number of zero symbols $k_0$. If $k_0 \geq h_2$, then it is the first write. If $h_2 > k_0 \geq h_3$, then it is the second write; and so on. If $h_t > k_0$, then it is the last write. Next, the information message is decoded from the number of zeros, the positions of the zeros, and the values of the nonzero entries (with erased symbols discarded in all but the first generation). ∎

Since the encoding and decoding for position modulation can be done with polynomial complexity (in $\log v_1$, ..., $\log v_t$), the resulting position modulation code has polynomial encoding and decoding complexity.

### A. Determining the Code Parameters

To find the coding parameters, we determine the $t$ numbers $h_1, \ldots, h_t$ in reverse order. First, we choose $h_t$ as the smallest number that satisfies condition (d). More specifically, we choose $h_t$ to be:

$$h_t = \left\lceil \frac{\log_2(v_t + 1)}{\log_2(2^m - 1)} \right\rceil. \tag{6}$$

Next, we choose $h_{t-1}$ as the smallest number greater than $h_t$ that satisfies condition (c); and so on. More specifically, for $1 < i < t$ we choose $h_i$ to be:

$$h_i = h_{i+1} + \delta_i, \tag{7}$$

$$\delta_i = \min\left\{\delta \,\middle|\, \sum_{k=1}^{\delta} \binom{h_{i+1} + \delta}{k} (2^m - 2)^k \geq v_i \right\} \tag{8}$$

We choose $h_1$ to be the smallest number that satisfies condition (b), i.e.,

$$h_1 = h_2 + \delta_1, \tag{9}$$

$$\delta_1 = \min\left\{\delta \,\middle|\, \sum_{k=0}^{\delta} \binom{h_2 + \delta}{k} (2^m - 1)^k \geq v_1 \right\} \tag{10}$$

**Lemma 1:** Consider given parameters $m, t$, and $v_1 = \ldots = v_t = v$. Let $\delta_i = h_i - h_{i+1}$. The above process (namely (6)-(10)) will output an increasing sequence $h_t, h_{t-1}, \ldots, h_1$ with nondecreasing increments, i.e.,

$$\delta_{t-1} \geq \delta_{t-2} \geq \ldots \geq \delta_1. \tag{11}$$

**Proof:** From (8), we have

$$\sum_{k=1}^{\delta_i} \binom{h_{i+1} + \delta_i}{k} (2^m - 2)^k \geq v. \tag{12}$$

Since $h_i > h_{i+1}$, this implies that

$$\sum_{k=1}^{\delta_i} \binom{h_i + \delta_i}{k} (2^m - 2)^k \geq v. \tag{13}$$

and

$$\sum_{k=0}^{\delta_i} \binom{h_i + \delta_i}{k} (2^m - 1)^k \geq v. \tag{14}$$

From (8) and (10) we know that $\delta_{i-1} \leq \delta_i$. ∎

*Example 1:* Consider a code that can write 56 bits of data 10 times using 278 wits, where a symbol is formed by $m = 2$ wits. The rate of this code is:

$$\frac{56 \times 10}{139 \times 2} = 2.01\ldots$$

In this case, the code paramters are:

$$h_{10} = 36,$$
$$h_9 = 51,$$
$$h_8 = 64,$$
$$h_7 = 76,$$
$$h_6 = 88,$$
$$h_5 = 99,$$
$$h_4 = 110,$$
$$h_3 = 120,$$
$$h_2 = 130,$$
$$h_1 = 139.$$

Note from (6)–(10) that if we want to obtain a code that writes 56 bits of data for $t' > 10$ times, then the first 9 numbers in the above list will remain the same.

## B. Performance Characterization

We next present a performance characterization of the position modulation code, by comparing it with a performance bound on womcodes in [18]. The original bound in [18] is for the $\langle v \rangle^t / n$ case; however, the argument can easily be extended to the $\langle v_1, \ldots, v_t \rangle / n$ case, as mentioned in [16]. The following lemma presents the bound for the $\langle v_1, \ldots, v_t \rangle / n$ case.

**Lemma 2 (Bound on womcodes [18]):**
Let $w(\langle v_1, \ldots, v_t \rangle)$ denote the least $n$ for which a $\langle v_1, \ldots, v_t \rangle / n$-womcode exists. Then

$$w(\langle v_1, \ldots, v_t \rangle) \geq Z_t(v_1, \ldots, v_t). \tag{15}$$

Here $Z_t(\ldots)$ is a $t$-variable function defined recursively as follows:

$$Z_t(v_1, \ldots, v_t) = Z_{t-1}(v_2, \ldots, v_t) + \delta(v_1, Z_{t-1}(v_2, \ldots, v_t)), \\ t \geq 1, \tag{16}$$

$$Z_0 = 0, \tag{17}$$

$$\delta(v, m) \triangleq \min\left\{\delta \left| \sum_{i=0}^{\delta} \binom{m+\delta}{i} \geq v \right.\right\} \tag{18}$$

**Proof:** The proof is by induction on $t$. The case $t = 0$ is trivial. Now consider any $\langle v_1, \ldots, v_t \rangle / n$ womcode for $t > 0$. For this code, after the first write, there must be at least $Z_{t-1}(v_2, \ldots, v_t)$ zeros. (If there are less than $Z_{t-1}(v_2, \ldots, v_t)$ zeros after the first write, it is not possible to accommodate the remaining $t-1$ writes since $w(\langle v_2, \ldots, v_t \rangle) \geq Z_{t-1}(v_2, \ldots, v_t)$.)

Therefore, the first write can only use codewords with at most $\kappa \triangleq n - Z_{t-1}(v_2, \ldots, v_t)$ ones. Since the first generation needs to represent a variable of size $v_1$, we must have:

$$\sum_{i=0}^{\kappa} \binom{n}{i} \geq v_1 \tag{19}$$

From the definition of $\delta(v, m)$, we see that

$$\kappa \geq \delta(v_1, Z_{t-1}(v_2, \ldots, v_t)). \tag{20}$$

Thus

$$n \geq Z_{t-1}(v_2, \ldots, v_t) + \delta(v_1, Z_{t-1}(v_2, \ldots, v_t)). \tag{21}$$

Since (21) holds for any $\langle v_1, \ldots, v_t \rangle / n$ womcode, $w(\langle v_1, \ldots, v_t \rangle) \geq Z_t(v_1, \ldots, v_t)$. ∎

**Theorem 2:** Given $v_1, \ldots, v_t$ with $v_i \geq 2$, consider $h_1, \ldots, h_t$ determined based on (6)–(10) for $m = 2$. Then

$$h_1 \leq Z_t(v_1, \ldots, v_t) \leq w(\langle v_1, \ldots, v_t \rangle). \tag{22}$$

Therefore, the position modulation code achieves a rate that is at least half the optimal rate.

**Proof:** With $m = 2$, we have

$$h_t = \left\lceil \frac{\log_2(v_t + 1)}{\log_2 3} \right\rceil \tag{23}$$

$$h_{t-1} = h_t + \delta_{t-1}, \tag{24}$$

$$\vdots$$

$$h_1 = h_2 + \delta_1, \tag{25}$$

where

$$\delta_{t-1} = \min\left\{\delta \left| \sum_{k=1}^{\delta} \binom{h_t + \delta}{k} 2^k \geq v_{t-1} \right.\right\}, \tag{26}$$

$$\delta_{t-2} = \min\left\{\delta \left| \sum_{k=1}^{\delta} \binom{h_{t-1} + \delta}{k} 2^k \geq v_{t-2} \right.\right\}, \tag{27}$$

$$\vdots$$

$$\delta_1 = \min\left\{\delta \left| \sum_{k=0}^{\delta} \binom{h_2 + \delta}{k} 3^k \geq v_1 \right.\right\} \tag{28}$$

By expanding the expression of $Z_t(v_1, \ldots, v_t)$, we have

$$Z_1(v_t) = \delta(v_t, 0) \tag{29}$$

$$Z_2(v_{t-1}, v_t) = Z_1(v_t) + \delta'_{t-1} \tag{30}$$

$$\vdots \tag{31}$$

$$Z_t(v_1, \ldots, v_t) = Z_{t-1}(v_2, \ldots, v_t) + \delta'_1, \tag{32}$$

where

$$\delta'_{t-1} \triangleq \delta(v_{t-1}, Z_1(v_t)) \tag{33}$$

$$\delta'_{t-2} \triangleq \delta(v_{t-2}, Z_2(v_{t-1}, v_t)) \tag{34}$$

$$\vdots$$

$$\delta'_1 \triangleq \delta(v_1, Z_{t-1}(v_2, \ldots, v_t)). \tag{35}$$

To simplify notations, we use the shorthand notation $Z_k$ to refer to $Z_k(v_{t-k+1}, \ldots, v_t)$. We now show via induction over $k$ that

$$h_{t-k+1} \leq Z_k. \tag{36}$$

The case $k = 1$ follows from the fact that for $v_t > 1$,

$$\delta(v_t, 0) = \lceil \log_2 v_t \rceil \geq \left\lceil \frac{\log_2(v_t + 1)}{\log_2 3} \right\rceil = h_t. \tag{37}$$

Suppose $h_{t-k+1} \leq Z_k$. We now show that $h_{t-k} \leq Z_{k+1}$. Note that $\delta'_{t-k} \geq 1$ and

$$\sum_{k=1}^{\delta'_{t-k}} \binom{Z_{k+1}}{k} 2^k \geq \sum_{k=0}^{\delta'_{t-k}} \binom{Z_{k+1}}{k} \geq v_{t-k} \tag{38}$$

$$\sum_{k=0}^{\delta'_{t-k}} \binom{Z_{k+1}}{k} 3^k \geq \sum_{k=0}^{\delta'_{t-k}} \binom{Z_{k+1}}{k} \geq v_{t-k}. \tag{39}$$

Since $Z_k \geq h_{t-k+1}$, $Z_k + \delta'_{t-k} - h_{t-k+1} \geq \delta'_{t-k}$. Thus

$$\sum_{k=1}^{Z_k+\delta'_{t-k}-h_{t-k+1}} \binom{Z_{k+1}}{k} 2^k \geq v_{t-k} \qquad (40)$$

$$\sum_{k=0}^{Z_k+\delta'_{t-k}-h_{t-k+1}} \binom{Z_{k+1}}{k} 3^k \geq v_{t-k}. \qquad (41)$$

Since $Z_{k+1} = Z_k + \delta'_{t-k} - h_{t-k+1} + h_{t-k+1}$, from (26)–(28) we see that

$$\delta_{t-k} \leq Z_k + \delta'_{t-k} - h_{t-k+1}, \qquad (42)$$

or equivalently,

$$h_{t-k} \leq Z_{k+1}. \qquad (43)$$

Thus, using induction we can show that $h_1 \leq Z_t(v_1, \ldots, v_t)$. ∎

## V. Performance Comparison

In this section we compare the performance of the position modulation codes with existing wom-codes. For the position modulation code, there are some cases where setting $m = 3$ gives slightly better performance than setting $m = 2$. For example, to write 56 bits twice, the position modulation code needs 98 wits with $m = 2$, and 96 wits with $m = 3$. However, setting $m = 2$ generally gives good performance among all choices of $m$. In the following, we use $m = 2$ for the position modulation code.

Table I gives the performance comparison of position modulation code (we use $v = 2^{56}$) and known low complexity $\langle v \rangle^t/n$ codes with the best rates, for $t = 2, \ldots, 10$. The known low complexity $\langle v \rangle^t/n$ codes with the best rates are given in the top row of the table. The position modulation codes and their rates are given in the bottom row of the table.

We now explain the codes in the top row from left to right. The $\langle 26 \rangle^2/7$ code is presented in [18] and this code is found via computer search, according to [18]. The $\langle 63 \rangle^3/12$ code is from James B. Saxe's construction of a $\langle 65, 81, 63 \rangle/12$ code, which was mentioned in [18]. There are two constructions of $\langle 7 \rangle^4/7$ codes. One is a cyclic womcode constructed by David Leavitt (mentioned in [18]) and the other is a womcode constructed from projective geometry by Frans Merkx [16]. The $\langle 11 \rangle^5/11$ womcode, designed by M. Beveraggi, is based on Steiner pentagonal systems, according to [3]. There are two constructions of the $\langle 16 \rangle^6/15$ code. One is the linear coset code given in [3] (Proposition 5 in [3]). The other is the linear code given in [18]. The $\langle 15 \rangle^7/15$ womcode is also constructed from projective geometry [16]. The next three codes are all obtained by combining the $\langle 15 \rangle^7/15$ womcode with some other small womcodes, since there are no known specific designs for $t = 8, 9, 10$ with rates higher than 1.62. This is done by using two observations given in [18]. First, by concatenating a $\langle v \rangle^{t_1}/n_1$ code and a $\langle v \rangle^{t_2}/n_2$ code side by side, we can have a $\langle v \rangle^{t_1+t_2}/(n_1+n_2)$ code (Information is represented by the modulo-$v$ sum of the values of the two subcodes). Second, by concatenating a $\langle v_1 \rangle^t/n_1$ code and a $\langle v_2 \rangle^t/n_2$ code side by side, we obtain a $\langle v_1 \cdot v_2 \rangle^t/(n_1+n_2)$ wom-code (Information is represented as the ordered pair of the two values). We obtain the $\langle 15 \rangle^8/19$ code by concatenating the $\langle 15 \rangle^7/15$ code with a $\langle 15 \rangle^1/4$ code (simply using the 4 bits to represent a 15-ary variable). We obtain the $\langle 15 \rangle^9/21$ code by concatenating the $\langle 15 \rangle^7/15$ code with a $\langle 15 \rangle^2/6$ code (based on the $\langle 16 \rangle^2/6$ code given in [18]). We obtain the $\langle 15 \rangle^{10}/24$ code by concatenating the $\langle 15 \rangle^7/15$ code with a $\langle 15 \rangle^3/9$ code, which is obtained by concatenating a $\langle 5 \rangle^3/5$ code and a $\langle 3 \rangle^3/4$ code. The $\langle 5 \rangle^3/5$ code is a cyclic code designed by David Klarner (see [18]) and the $\langle 3 \rangle^3/4$ code is given in [4].

For each $t \in \{2, \ldots, 10\}$, the best rate is shown in boldface. It is seen that the position modulation code offers higher rates than state-of-the-art solutions for $t = 5, 6, 8, 9, 10$.

For larger values of $t$, we compare the position modulation code with three existing general code designs that produce classes of codes. In [18], Rivest and Shamir presented a $\langle v \rangle^{1+v/4}/(v-1)$ linear code. The rate of this code is less than 2 for $t \leq 50$. In [4], Fiat and Shamir presented a $\langle 3 \rangle^t/(t+1)$ code that works for arbitrary $t$. The rate of this code is always less than 1.59. In [3], Cohen *et al.* presented a $\langle 2^r \rangle^{2^{r-2}+2}/(2^r-1)$ code that works for $r \geq 4$; according to the paper, the encoding process is NP-hard. Figure 3 shows the rates of these wom-coding schemes, where the position modulation code is for $m = 2$ and $v = 2^{32}$. It is seen that the position modulation code achieves higher rates.

## VI. Conclusion

We presented a method for constructing wom codes with low encoding and decoding complexity and high rates, which works for general problem configurations. The proposed method is called *position modulation code*, as it is built upon a simple and yet fundamental observation, that positions of zeros in a binary array can be used to encode information. The position modulation code applies position modulation to symbols formed by multiple cells and performs position modulation in all writes except the last one, by using a "soft" erasing operation to reset the state between writes.

It is proven that the position modulation code achieves a rate that is at least half of the optimal rate. Furthermore, the proposed position modulation codes is seen to have superior rates compared to existing wom codes. The low complexity and high rate of the position modulation code make it a promising candidate for practical applications.

In addition, as a family of codes that accommodate general parameter configurations, the position modulation code can be readily used in related rewriting schemes. For example, the trajectory code [12] for a generalized rewriting model can use the position modulation code as a subcode for better performance.

TABLE I
PERFORMANCE COMPARISON OF POSITION MODULATION CODE AND KNOWN $\langle v \rangle^t/n$ CODES

| | $t=2$ | $t=3$ | $t=4$ | $t=5$ | $t=6$ | $t=7$ | $t=8$ | $t=9$ | $t=10$ |
|---|---|---|---|---|---|---|---|---|---|
| Known $\langle v \rangle^t/n$ codes | $\langle 26 \rangle^2/7$ **R = 1.34** | $\langle 63 \rangle^3/12$ **R = 1.49** | $\langle 7 \rangle^4/7$ **R = 1.60** | $\langle 11 \rangle^5/11$ $R = 1.57$ | $\langle 16 \rangle^6/15$ $R = 1.60$ | $\langle 15 \rangle^7/15$ **R = 1.82** | $\langle 15 \rangle^8/19$ $R = 1.65$ | $\langle 15 \rangle^9/21$ $R = 1.67$ | $\langle 15 \rangle^{10}/24$ $R = 1.63$ |
| Position modulation code | $\langle 2^{56} \rangle^2/98$ $R = 1.14$ | $\langle 2^{56} \rangle^3/124$ $R = 1.35$ | $\langle 2^{56} \rangle^4/150$ $R = 1.49$ | $\langle 2^{56} \rangle^5/172$ **R = 1.63** | $\langle 2^{56} \rangle^6/196$ **R = 1.71** | $\langle 2^{56} \rangle^7/216$ $R = 1.81$ | $\langle 2^{56} \rangle^8/238$ **R = 1.88** | $\langle 2^{56} \rangle^9/258$ **R = 1.95** | $\langle 2^{56} \rangle^{10}/278$ **R = 2.01** |

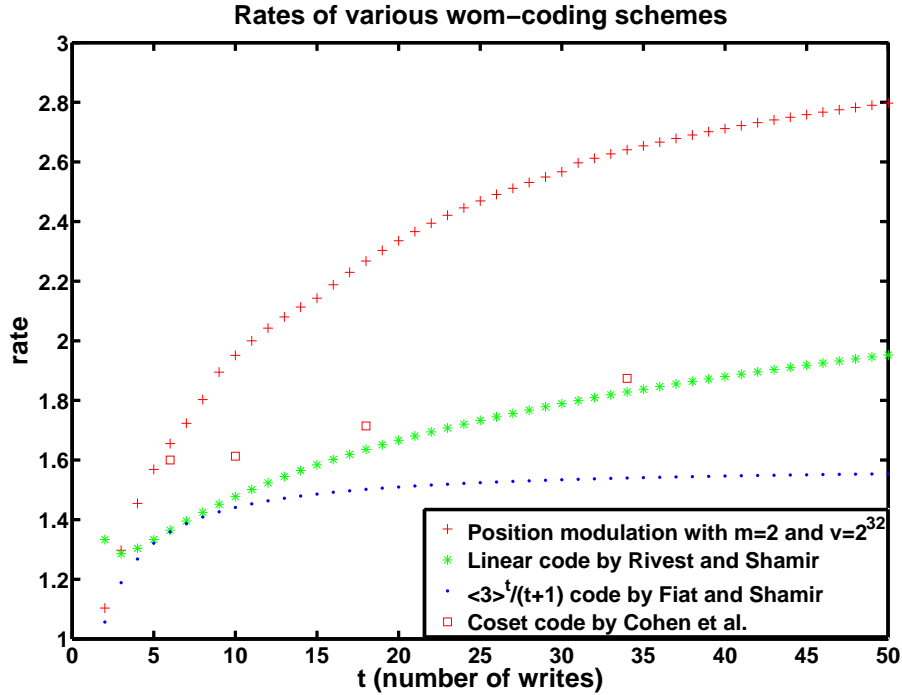

Fig. 3. Rates for various wom-coding schemes.